# The B-Skip-List: A Simpler Uniquely Represented Alternative to B-Trees

Daniel Golovin[*]

**Abstract**

In previous work, the author introduced the *B-treap*, a uniquely represented B-tree analogue, and proved strong performance guarantees for it. However, the B-treap maintains complex invariants and is very complex to implement. In this paper we introduce the *B-skip-list,* which has most of the guarantees of the B-treap, but is vastly simpler and easier to implement. Like the B-treap, the B-skip-list may be used to construct *strongly history-independent* index structures and filesystems; such constructions reveal no information about the historical sequence of operations that led to the current logical state. For example, a uniquely represented filesystem would support the deletion of a file in a way that, in a strong information-theoretic sense, provably removes all evidence that the file ever existed.

Like the B-tree, the B-skip-list has depth $O(\log_B n)$ where $B$ is the block transfer size of the external memory, uses linear space with high probability, and supports efficient one-dimensional range queries.

## 1 Introduction

Uniquely represented data structures represent each logical state with a unique machine state. Because they have canonical representations, such data structures are *strongly history-independent*, and thus reveal no information about the historical sequence of operations that led to the current logical state. This has useful applications in cases where strict privacy or high security is required, because most computer applications store a significant amount of information that is hidden from the application interface—sometimes intentionally but more often not. This information might consist of data left behind in memory or disk, but can also consist of much more subtle variations in the state of a structure due to previous actions or the ordering of the actions. Such unintentionally stored information might be leaked to the wrong people with disastrous consequences. For example there are many embarrassing stories about improperly redacted PDFs being released to the public. Chapter 1 of [9] gives examples of improper redactions involving leaks by the Central Intelligence Agency in 2000 revealing information about its role in the 1953 overthrow of the Iranian Government, by the United States Army in 2005 about the accidental shooting of Italian intelligence agent Nicola Calipari in Iraq, and by the Fédération Internationale de l'Automobile in 2007 of confidential and valuable technical information on the McLaren and Ferrari racing cars. A more recent example involved the Transportation Security Administration revealing classified information about airport passenger screening practices in 2009.

To address the concern of releasing historical and potentially private information various notions of *history independence* have been invented along with data structures that support these notions [12, 14, 11, 5, 1]. Unique representation had been studied even earlier [18, 19, 2], in the context of theoretical investigations into the need for redundancy in efficient data structures.

[*]Center for the Mathematics of Information, California Institute of Technology, Pasadena, CA 91125. Email: *dgolovin@caltech.edu* .

**Unique Representation on a RAM.** There has been much recent progress on efficient uniquely represented data structures in the RAM model. Blelloch and Golovin [3] described a uniquely represented hash table for the RAM model supporting insertion, deletion and queries in expected constant time, using linear space and only $O(\log n)$ random bits. They also provided a perfect hashing scheme that allows for $O(1)$ worst-case queries, and efficient uniquely represented data structures for ordered dictionaries and the order maintenance problem. Naor *et al.* [13] developed a second uniquely represented dynamic perfect hash table supporting deletions, based on cuckoo hashing. Blelloch *et al.* [4] developed efficient uniquely represented data structures for some common data structures in computational geometry. Several other results, as well as a more comprehensive discussion of uniquely represented data structures, may be found in the author's doctoral thesis [9].

**Unique Representation in External Memory Models.** Progress in the RAM model suggested that it might be possible to construct efficient uniquely represented data structures for organizing information on disk, i.e., for *external memory* (EM) models of computation [21] which account for the fact that modern computers have a memory hierarchy in which external memory (e.g., a disk drive) is several orders of magnitude slower than internal memory (e.g., DRAM). For background on the extensive body of work on conventional data structures in EM models, we refer the interested reader to the excellent book by Vitter [21]. Within this body of work, *extendible hash tables* and the *B-tree* and its variants (e.g., the $B^+$-tree and the $B^*$-tree) play a prominent role. It is worth noting here that the extendible hashing construction of Fagin *et al.* [8] is almost uniquely represented, and in fact can be made uniquely represented with some minor modifications, the most significant of which is to use a uniquely represented hash table [3, 13] to layout blocks on disk. However, in this paper we focus on uniquely represented B-tree analogs, which can support efficient one-dimensional range queries. The first such uniquely represented B-tree analog was the B-treap developed by the author [10]. It supports the following operations.

- insert($x$): insert element $x$.
- delete($x$): delete element $x$.
- lookup($x$): determine if $x$ is present, and if so, return a pointer to it.
- range-query($x, y$): return all elements between $x$ and $y$ stored in the data structure.

It is easy to associate auxiliary data with the elements, though for simplicity of exposition we will assume there is no auxiliary data being stored.

The following result, proved in [10] indicates that the B-treap has performance comparable to B-trees, up to constant factors. Empirical results in [10] suggest that these constants are extremely modest in practice (e.g., less than 1.5 factor increase in depth).

**Theorem 1.1 ([10])** *There exists a uniquely represented B-treap that stores elements of a fixed size, such that if the B-treap contains $n$ elements and the block transfer size $B = \Omega\left((\ln(n))^{1/(1-\epsilon)}\right)$ for some $\epsilon > 0$, then* lookup, insert, *and* delete *each touch at most $O(\frac{1}{\epsilon}\log_B(n))$ B-treap nodes in expectation, and* range-query *touches at most $O(\frac{1}{\epsilon}\log_B(n) + k/B)$ B-treap nodes in expectation where $k$ is the size of the output. Furthermore, if $B = O(n^{\frac{1}{2}-\delta})$ for some $\delta > 0$, then with high probability the B-treap has depth $O(\frac{1}{\epsilon}\log_B(n))$ and requires only linear space to store.*

The B-treap thus performs the functions of a B-tree in a uniquely represented manner with remarkably low overhead. However, as a practical matter the B-treap is an extremely complicated data structure and is difficult to implement. For this reason we introduce the B-skip-list, which has nearly all of the same theoretical guarantees, but is conceptually simpler and far easier to implement. As the name suggests, the B-skip-list is based on the skip-lists of Pugh [17]. We prove the following guarantees for the B-skip-list.

**Theorem 1.2** *There exists a uniquely represented B-skip-list that stores elements of a fixed size, such that if the B-skip-list contains $n$ elements and the block transfer size is $B$, then* lookup, insert, *and* delete *each touch at most $O(\log_B(n))$ external memory blocks in expectation, and* range-query *touches at most $O(\log_B(n) + k/B)$ external memory blocks in expectation where $k$ is the size of the output. Furthermore, with high probability the B-skip-list requires only linear space to store.*

## 2 Preliminaries

**Terminology & Basic Notation.** We let $\mathcal{X}$ denote the (ordered) universe of elements that we may store. Since the B-skip-list will store duplicates of some of the elements, we say that the B-skip-list is comprised of *nodes* which are *labelled* with the element that they store. Nodes may also contain *abstract pointers* to other nodes, as described in Section 2.2.1. For $n \in \mathbb{Z}$, let $[n]$ denote $\{1, 2, \ldots, n\}$.

### 2.1 The External Memory Model

We use a variant of the parallel disk model of Vitter [21] with one processor and one disk, which measures performance in terms of disk I/Os. Internal memory is modeled as a 1-D array of data items, as in a RAM. External memory is modeled as a large 1-D array of *blocks* of data items. A block is a sequence of $B$ data items, where $B$ is a parameter called the *block transfer size*. The external memory can read (or write) a single block of data items to (or from) internal memory during a single I/O. Other parameters include the problem size, $n$, and the internal memory size $m$, both measured in units of data items. We will assume $m = \omega(B)$, so that any constant number of blocks can be stored in internal memory.

### 2.2 Uniquely Represented Dynamic Memory Allocation

In this section we show how to perform uniquely represented memory allocation, which allows us to reduce the problem of constructing uniquely represented data structures to the problem of constructing data structures with a canonical pointer structure.

#### 2.2.1 Memory Allocation with Fixed Length Keys

Uniquely represented hash tables [3, 13] can be used as the basis for a uniquely represented memory allocator. Intuitively, if the nodes of a pointer structure can be labeled with distinct hashable labels in a uniquely represented (i.e., strongly history independent) manner, and the pointer structure itself is uniquely represented in a pointer based model of computation, then these hash tables provide a way of mapping the pointer structure into a one dimensional memory array while preserving unique representation. Pointers are replaced by labels, and pointer dereferencing is replaced by hash table lookups. We call such labels *abstract pointers*. We will assume that the data items have distinct hashable labels, which can be used to generate distinct hashable labels for the nodes of the B-skip-list in a uniquely represented manner.

Given these labels, and a data structure composed of nodes and abstract pointers with a uniquely represented pointer structure, we can hash the nodes of the data structure into external memory. If we use distinct random bits for the hash table and everything else, this inflates number of the expected I/Os by a constant factor. (For B-skip-lists, uniquely represented dynamic perfect hash tables [3, 13] may be an attractive option, since in expectation most of the I/Os will involve reads, even in the case of insertions and deletions.)

### 2.2.2 Memory Allocation with Variable Length Keys

In our external memory model, we can obtain significantly better performance if we allocate large data objects in consecutive regions of the external memory, rather than breaking them up and hashing the pieces to various pseudorandom locations. Therefore we would like to perform efficient memory allocation with variable length keys. In addition to the undesirable solution of breaking large objects into smaller pieces and using a uniquely represented hash table $H$ to allocate each piece directly using $H$, there are at least two other options. First, we could use $H$ as a virtual memory scheme which allocates fixed sized regions of consecutive blocks to each large data object, and then using another memory allocator to store the object as though its blocks are the entire memory. A second method, which is more elegant and arguably more suitable for external memory models, is to simply store variable sized objects directly in $H$, as intervals in the memory array. Recently Thorup [20] analyzed the running time of such a variant of linear probing. He showed that using sufficiently random hash functions, this variant supports operations in time linear in the string length plus a measure of the variance of the string lengths, namely the ratio of the second moment of the string lengths to their mean; for the cases we are interested in the latter will be constant. This variant stores a set of strings $S$ (each of which is assumed to terminate with an end-of-string character) from a universe of strings $\mathcal{S}$ such that each string $s$

- is stored in a contiguous block of hash table slots
- is preceded in memory by either an end-of-string character or an empty slot
- is stored such that its first character lies between $h(s)$ and the first empty slot after $h(s)$ (i.e., the empty slot $y$ minimizing $y - h(s) \mod p$ if there are $p$ slots) in the table storing keys $S \setminus \{s\}$, inclusive.

It is straightforward to see that the uniquely represented hash table of Blelloch and Golovin [3] can be easily modified to store variable length objects in this way. To prove unique representation, we can reduce this case to the case with fixed length keys handled in [3] as follows. Treat each string $s = s_1 \cdots s_k$ of $k$ characters as $k$ keys $\{s_1, \ldots, s_k\}$ such that each $s_i$ hashes to the same location and each slot prefers $s_i$ slightly to $s_{i+1}$ for all $i$, but for all distinct strings $s$ and $s'$, each slot either prefers $s_i$ to $s'_j$ for all $i$ and $j$ or each slot prefers $s'_j$ to $s_i$ for all $i$ and $j$. For the required slot preferences over the keys, we may use lexicographic ordering for the strings, and derive the relative slot preferences for the characters of each string as described above. Note that we describe this reduction only for purposes of analysis and to provide insight; we do not suggest implementing the memory allocator in this manner, i.e., by treating each string as a sequence of keys. The problem has structure that allows us to work with the strings directly. The running time for the operations is thus dominated by the displacement[1] and the length of the input string. Hence Thorup's bounds in Theorem 6.1 of [20] apply to our modified table as well. Before we describe the technical result which follows from this observation, we recall the definition of $k$-universal hash functions [22].

**Definition 2.1 ($k$-Universal Hash Functions)** *A $k$-universal hash function from $X$ to $Y$ is a random function $h : X \to Y$ drawn from some distribution such that the images of any $k$ keys under $h$ are distributed uniformly at random. That is, for all distinct $x_1, \ldots, x_k \in X$ and for all $y_1, \ldots, y_k \in Y$, $\mathbf{Pr}_h \left[ \forall i \in [k] . h(x_i) = y_i \right] = 1/|Y|^k$.*

We are now ready to state our result.

**Theorem 2.2** *Let $S \subset \mathcal{S}$ be a set of variable length strings from universe $\mathcal{S}$. Let $\mathrm{length}(s)$ denote the length of string $s$, including an end-of-string character. Fix a parameter $p$ such that $\sum_{s \in S} \mathrm{length}(s) \leq \alpha p$, for load $\alpha < 0.9$. Let $h_1$ be a 1-universal hash function from $\mathcal{S}$ to $[p]$ and let $h_2$ be a 5-universal hash function from $[p]$ to $[p]$. Given a hash table $H$ and $s_0 \in \mathcal{S}$, let $\delta_H(s_0, S, h_1, h_2)$ denote the displacement of $s_0$ in $H$ if* $H$ is currently storing $S$ and using $h_2 \circ h_1$ as its hash function from $\mathcal{S}$ to $[p]$. Then there exists a uniquely represented hash table $H$ such that for all $s_0 \in \mathcal{S}$,

$$\mathbf{E}[\delta_H(s_0, S, h_1, h_2)] \leq \text{length}(s_0) + O\left(\alpha^{1/3} \frac{\sum_{s \in S} \text{length}^2(s)}{\sum_{s \in S} \text{length}(s)}\right)$$

---

[1] Recall that in a linear probing hash table, the *displacement* of a key $s$ is the number of slots $s$ must unsuccessfully probe before finding an empty slot. In other words, if $s$ is inserted into slot $y$ of a hash table with slots $\{0, 1, 2, \ldots, p-1\}$, its displacement is $(y - s) \mod p$.

*and for all $s_0 \notin S$,*

$$\mathbf{E}[\delta_H(s_0, S, h_1, h_2)] = O\left(\alpha \frac{\sum_{s \in S} \text{length}^2(s)}{\sum_{s \in S} \text{length}(s)}\right)$$

*where the expectation is taken over $(h_1, h_2)$.*

Theorem 2.2 only gives bounds on the displacement of $s_0$, rather than the total amount of work needed to, say, insert $s_0$. The latter quantity is roughly equal to the displacement, plus the length of $s_0$, plus the number of characters we need to displace (i.e., move forward) to make room for $s_0$. However, this latter quantity is equal to the displacement of a new key $s_0'$ immediately after $s_0$ is inserted, assuming $s_0'$ hashes to same location as $s_0$. Thus we can convert the bounds on displacement in Theorem 2.2 into bounds on running time with only a slight increase in the bound due to $s_0$. It is worth noting that in our application, $\frac{\sum_{s \in S} \text{length}^2(s)}{\sum_{s \in S} \text{length}(s)}$ will be constant with high probability, and Theorem 2.2 will be used to prove that with the right hash functions our variable length key memory allocator will take expected time linear in the key length to insert, delete, or lookup a key.

## 3 B-Skip-Lists

### 3.1 Review of Skip Lists

Skip lists were developed by William Pugh [17] as a simple randomized alternative to balanced binary search trees. Let $X$ be the set of elements we wish to store. The idea behind skip lists is assign a random positive integer *level* to each element $x \in X$, such that $\mathbf{Pr}[\text{level}(x) = k]$ decreases exponentially in $k$, and for each $k$ such that some element has $\text{level}(x) \geq k$, maintain a list $L_k$ of all elements $x$ with $\text{level}(x) \geq k$. We also assume there is a special element front in every list that is smaller than all other elements, so as to be at the front of every list. Note each list $L_k = \langle x_1, x_2, \ldots, x_t \rangle$ partitions the elements $X$ into contiguous subsets

$$\{\{x : x_i \leq x < x_{i+1}\} : 1 \leq i < t\} \cup \{\{x : x_t \leq x\}\}$$

and the partition induced by $L_{k-1}$ refines[2] the partition induced by $L_k$. Additionally, we maintain pointers from each element $x \in L_k$ to $x \in L_{k-1}$. Let $x_i$ be the *representative* of $\{x : x_i \leq x < x_{i+1}\}$. When searching for an element $x$, we start at the highest level list $L_k$ and search for the representative $y_k$ of the level $k$ partition set containing $x$, then starting from $y_k$ in $L_{k-1}$ search for the representative $y_{k-1}$ of the level $k-1$ partition set containing $x$, and so on until reaching the representative $y_1 = x$ of the level one partition set containing $x$. See Figure 1.

The original skip list paper [17] sets element levels by repeatedly flipping a biased coin up to $(\beta - 1)$ times, for some parameter $\beta$ which bounds the maximum allowable level. The level of the element is then set to one plus the number of coin flips to obtain an outcome of 'heads'. If all $(\beta - 1)$ coin flips turn up 'tails,' then the level is set to $\beta$. The distribution over levels is thus parametrized by the probability of tails $q \in (0, 1)$ and a bound $\beta$ on the maximum allowable level. Specifically, it is

$$\mathbf{Pr}[\text{level}(x) = k] = \begin{cases} q^{(k-1)}(1-q) & \text{if } 1 \leq k < \beta \\ q^{(\beta-1)} & \text{if } k = \beta \\ 0 & \text{if } k < 1 \text{ or } k > \beta \end{cases} \quad (3.1)$$

---

[2] A partition $\{A_1, \ldots, A_r\}$ of $X$ is said to refine a partition $\{B_1, \ldots, B_r\}$ of $X$ if for all $A_i$ there is a $B_j$ such that $A_i \subseteq B_j$.

Using this distribution with $q = 1/2$ and $\beta = \lceil \log_2(n) \rceil$ to sample the levels independently for each element, lookups, insertions, deletions, splits, and joins can be implemented in $O(\log n)$ time [17, 16].

Inspecting Figure 1, it is fairly easy to see that the pointer structure of the skip list is uniquely represented if the levels are generated using a hash function. We require that $\mathrm{level}(\cdot)$ be a deterministic function (which may be drawn at random from a suitable hash family), so that the levels are not influenced by what other elements are present, and so they are consistent if an element is deleted and later reinserted.

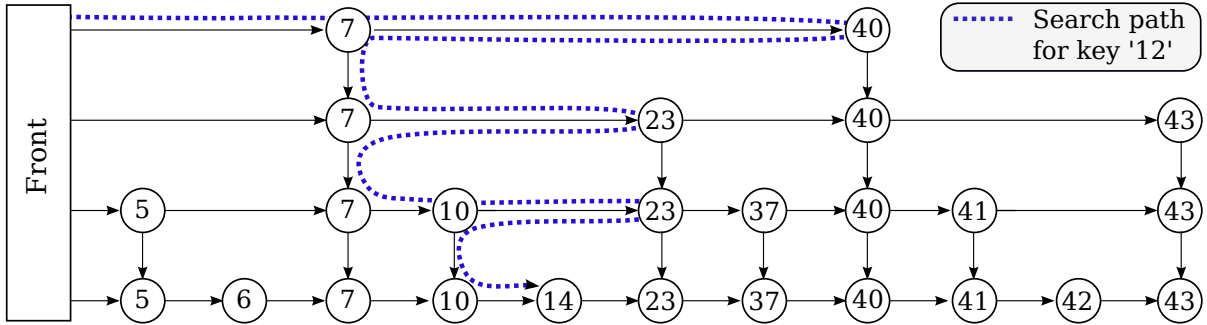

Figure 1: A skip list, with the search path for key 12 illustrated.

## 3.2 The B-Skip-List

We suppose we have a capacity $N$ bounding the maximum size (i.e., number of nodes) of the B-skip-list. Fix a parameter $\gamma$, whose purpose will be explained below. In practice $\gamma$ will typically be a small multiple of $B$. The B-skip-list is a "blocked version" of a uniquely represented skip-list on $n$ nodes, where the levels are chosen from the distribution in equation 3.1 with $q = 1/\gamma$ and $\beta = \lceil \log_\gamma(N) \rceil + 2$. To maintain unique representation, the distribution should be sampled using a hash function of the elements, so that if an element is deleted and then reinserted it has the same level. We refer the interested reader to Section 5.1 of [9] for further details on the construction of uniquely represented skip lists.

**The B-Skip-List Partition Invariant.** The "blocking" of the uniquely represented skip list is rather simple. Recall $L_k$ is the list of elements at level $k$ or higher. Let $L_k = \left\langle x_1^k, x_2^k, \ldots, x_{|L_k|}^k \right\rangle$. For each level $k$, we partition $L_k$ as

$$\{\{x : x \in L_k, \ x_i^{k+1} \leq x < x_{i+1}^{k+1}\} : 1 \leq i < |L_{k+1}|\} \cup \{\{x : x \in L_k, \ x_{|L_{k+1}|}^{k+1} \leq x\}\}$$

Refer to Figure 2 for an example. Each of these partitions will be a key (in the memory allocator) containing a list of skip list nodes. A list $\sigma$ of $|\sigma|$ skip list nodes is then stored across at most $\lceil |\sigma|/B \rceil + 1$ contiguous blocks of external memory. We allow $\sigma$ to be stored starting in the middle of a block. Note that the hash table underlying the memory allocator does not work over an array of blocks, but rather an array of finer grained subdivisions of blocks, e.g., over chunks of memory large enough to store a B-skip-list node. We say $\sigma$ *spans* the blocks in which it is stored. In the terminology of Section 2.2.2, $\sigma$ is a string of length $|\sigma|$. We call these constraints on how the B-skip-list is stored the *partition invariant*. The label to be used to hash a partition into external memory for purposes of memory allocation can be constructed from the minimum skip list node in the partition, and the level of the partition. Operations on the B-skip-list can proceed almost exactly as with the regular skip-list, with two differences. The first difference is that pointers between partitions must be replaced with abstract pointers (i.e., labels), and pointer dereferencing between partitions must be replaced with hash lookups, as discussed in Section 2.2 and in more detail in [3, 9]. The second difference is that the data structure must preform some more work than a standard skip list, in order to maintain the partitions and the invariants on how they are stored.

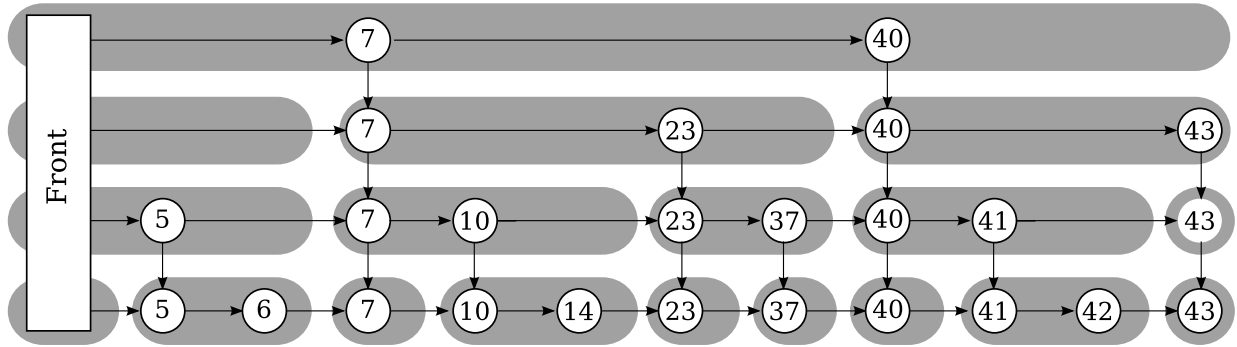

Figure 2: A skip list partitioned into smaller lists. Each partition corresponds to a gray region. Let $(x, k)$ denote the node labeled $x$ in list $L_k$. Then, e.g., $\{(7, 2), (10, 2)\}$ is a partition because it consists of an element of level greater than two, namely, 7, and all elements between it and the next such element, 23.

**The Tuning Parameter $\gamma$.** We can now describe the role of $\gamma$. The levels are geometric random variables with parameter $(1 - 1/\gamma)$ truncated at $\beta$. The size of a partition set is a random variable that is also described by a geometric random variable with parameter $1/\gamma$ and support $[n]$. As such the latter has expectation of roughly $\gamma$. The parameter $\gamma$ thus controls the expected size of a partition set, and for this reason $\gamma = \Theta(B)$ is appropriate in most applications. Since these partition sets are stored as strings over "characters" large enough to represent a B-skip-list node, the sizes of the partitions, and particularly how many blocks are required to store them, play an important role in the performance of the B-skip-list. Too high, and the insertion of an element may often cause large partition sets to be moved on disk during memory allocation, resulting in poor performance. Too low, and searches will take longer due to the increased height, because as $\gamma$ decreases the height $\lceil \log_\gamma N \rceil + 2$ increases.

**Maintaining the Partitions.** When an element $y$ with $\text{level}(y) \geq 2$ is inserted, it splits existing partitions in two. Specifically, if $(x, k)$ denotes the node labeled $x$ in list $L_k$, and $\langle (x_1, k), (x_2, k), \ldots, (x_t, k) \rangle$ is a partition before $y$ is inserted, and $\text{level}(y) > k$ and $x_i < y < x_{i+1}$ in element order for some $i$, then inserting $y$ splits the partition into $\langle (x_1, k), (x_2, k), \ldots, (x_i, k) \rangle$ and $\langle (y, k), (x_{i+1}, k), (x_{i+2}, k), \ldots, (x_t, k) \rangle$. Note that inserting $y$ causes one such split for each level less than $\text{level}(y)$. Each such split involves deleting the old partition, generating the new partition lists and memory allocating them.

Similarly, when an element $y$ with $\text{level}(y) \geq 2$ is deleted, its deletion causes $\text{level}(y) - 1$ existing partitions to merge, in effect appending the two lists of nodes together. That is, if there are partitions $\langle (x_1, k), (x_2, k), \ldots, (x_i, k) \rangle$ and $\langle (y, k), (x_{i+1}, k), (x_{i+2}, k), \ldots, (x_t, k) \rangle$, where $x_i$ immediately precedes $y$ in $L_k$, then deleting $y$ requires us to merge the two partitions into $\langle (x_1, k), (x_2, k), \ldots, (x_t, k) \rangle$. To find all such partitions, it suffices to search for the predecessor of $y$ in the B-skip-list. Thus when deleting $y$, we should actually search for its immediate predecessor $x$, and use the lowest level instance of $x$, i.e., $(x, 1)$, to find $y$. The path from the highest level instance of the front element to the predecessor $(x, 1)$ will then contain elements in all partition sets with elements less than $y$ that need to be merged during the its deletion.

These can easily be used to the find all the partition sets with elements greater than $y$ hat need to be merged during its deletion, simply by following the skip-list abstract pointers.

## 4 Analysis of the B-Skip-List

We analyze the B-skip-list using the number of I/Os as the cost of an operation. This cost model is common in external memory data structures and is justified insofar as disk I/Os are several orders of magnitude slower than main memory operations on modern hardware. To analyze the performance of the B-skip-list in this model, we need to bound its depth (i.e., the number of levels in it), understand the distribution on partition set size, and understand how the partition set size distribution impacts the time to perform the hash table operations involved in memory allocation in external memory.

For purposes of analysis, we will assume that $\{\text{level}(x) : x \in \mathcal{X}\}$ are independent, where $\mathcal{X}$ is the universe of elements. To prove similar theoretical guarantees without this assumption, note that there exist sophisticated, efficiently computable hash functions that are $N$-wise independent with high probability that if used to generate $\{\text{level}(x) : x \in \mathcal{X}\}$ ensure our results will hold with high probability [15, 7].

**Bounding the Depth**

We begin with the simple fact that, by construction, the depth is at most $\beta = \lceil \log_\gamma(N) \rceil + 2$. Recall $q = 1/\gamma$. For $n << N$, we can bound the depth as the maximum level of any element via the union bound. That is, $\mathbf{Pr}[\text{level}(x) \geq k] \leq q^{k-1}$, which implies $\mathbf{Pr}[\exists x, \text{level}(x) \geq k] \leq n \cdot q^{k-1}$, and so $\mathbf{Pr}\big[\text{depth} \geq (c+1)\log_\gamma(n)\big] \leq n^{-c}$.

**The Distribution on Partition Sizes**

We prove the following exponential tail bound on the distribution in partition set sizes.

**Lemma 4.1** *If $\gamma \geq 2$ then for each partition set $S$, $\mathbf{Pr}[|S| \geq \lambda] \leq \exp(-\lambda/\gamma)$.*

**Proof:** First we consider a partition set $S$ containing nodes at level $k < \beta$, where $\beta$ is the maximum possible level. Let $(x, k)$ denote the level $k$ instance of $x$. Let $(y, k)$ be the smallest node in $S$, by element order (i.e., the order in which $(y, k) < (x, k)$ if $y < x$). Consider the elements $x \in L_k$ greater than or equal to $y$. Each has $\text{level}(x) \geq k$, and if $\text{level}(x) > k$, then $(x, k)$ and everything greater than it in level $k$ are not in $S$. Straightforward calculation reveals $\mathbf{Pr}[\text{level}(x) > k \mid \text{level}(x) \geq k] = q$. Alternately, this fact is readily apparent if we note that each $\text{level}(x)$ can be generated by repeatedly flipping a coin with bias $(1-q)$ until it comes up heads, truncating after $\beta - 1$ flips if necessary. Since the levels are independent by assumption, and $|S| \geq \lambda$ implies $(y, k)$ and the succeeding $\lambda - 1$ nodes in $\text{level}(k)$ have not been assigned any level above $k$,

$$\mathbf{Pr}[|S| \geq \lambda] \leq (1-q)^\lambda \leq \exp(-q\lambda) = \exp(-\lambda/\gamma)$$

where we have used $1 - q \leq e^{-q}$ for all $q \in \mathbb{R}$.

It remains to prove the bound for the unique partition set $S_\beta$ at level $\beta = \lceil \log_\gamma(N) \rceil + 2$. For this, we note that $\mathbf{Pr}[x \in S] = q^{\beta-1}$, and given the events $\{x \in S\}$ are independent, we can write

$$\mathbf{Pr}[|S_\beta| \geq \lambda] \leq \mathbf{Pr}[\text{there exist } \lambda \text{ elements with level } \beta]$$
$$\leq \binom{n}{\lambda} q^{\lambda(\beta-1)}$$
$$\leq n^\lambda (1/\gamma)^{\lambda(\lceil \log_\gamma(N) \rceil + 1)}$$
$$\leq (n/N)^\lambda \cdot \gamma^{-\lambda}$$
$$\leq \gamma^{-\lambda} = \exp(-\lambda \ln(\gamma))$$

Assuming $\gamma \geq 2$, we see $\ln(\gamma) > 1/\gamma$ and thus we obtain the desired bound of $\mathbf{Pr}[|S_\beta| \geq \lambda\gamma] \leq \exp(-\lambda/\gamma)$ for this set as well. ∎

**The Running Time of Memory Allocation**

Lemma 4.1 in conjunction with Theorem 2.2 yields the following performance bound for the memory allocator of Section 2.2.2.

**Lemma 4.2** *If $\gamma = O(B)$, then the memory allocator operations of inserting, deleting, or looking up a partition set $S$ in external memory require only $O(|S|/B + 1)$ I/Os in expectation.*

**Proof:** Let $\mathcal{C}$ be the collection of partition sets. Consider the memory allocator as operating over strings representing partition sets in which the characters are B-skip-list nodes. Then by Theorem 2.2 the running time to insert, delete, or lookup $S \in \mathcal{C}$ will be $O(|S| + \mathbf{E}\left[\frac{\sum_{S \in \mathcal{C}} |S|^2}{\sum_{S \in \mathcal{C}} |S|}\right])$ in expectation. Since this is a bound on the displacement after $S$ is inserted, it implies that the number of external memory I/Os will be at most $1/B$ times this quantity, plus one additional I/Os to deal with the consideration that the partitions may not be block aligned, but may start in the middle of a block. We now bound $\mathbf{E}\left[\frac{\sum_{S \in \mathcal{C}} |S|^2}{\sum_{S \in \mathcal{C}} |S|}\right]$. First, note that for all positive real $a_1, \ldots, a_m$, we have $\sum_i a_i^2 \leq (\sum_i a_i)^2$. Thus

$$\frac{\sum_{S \in \mathcal{C}} |S|^2}{\sum_{S \in \mathcal{C}} |S|} \leq \sum_{S \in \mathcal{C}} |S|$$

and so we take expectations and bound $\mathbf{E}\left[\frac{\sum_{S \in \mathcal{C}} |S|^2}{\sum_{S \in \mathcal{C}} |S|}\right]$ by $\mathbf{E}\left[\sum_{S \in \mathcal{C}} |S|\right]$. Using Lemma 4.1 and $\mathbf{E}[X] = \sum_{x \geq 1} \mathbf{Pr}[X \geq x]$ for nonnegative integer valued random variables $X$, we obtain

$$\mathbf{E}[|S|] \leq \sum_{x \geq 1} e^{-x/\gamma} = \frac{e^{-1/\gamma}}{1 - e^{-1/\gamma}} = \Theta(\gamma)$$

Supposing $\gamma = O(B)$, the result is that

$$\mathbf{E}\left[\frac{\sum_{S \in \mathcal{C}} |S|^2}{\sum_{S \in \mathcal{C}} |S|}\right] = O(B)$$

and so the contribution of external memory I/Os by the $\frac{\sum_{S \in \mathcal{C}} |S|^2}{\sum_{S \in \mathcal{C}} |S|}$ portion of the displacement bound is constant. Hence the expected number of external memory I/Os will be $O(|S|/B + 1)$. ∎

**Running Time Analysis of the B-Skip-List**

We are almost ready to analyze the running time of the B-skip-list, but first we must introduce some notation and prove a lemma. Consider a conventional skip-list $\mathrm{lookup}(y)$ operation, and suppose the search path is $\langle(x_1, k_1), \ldots, (x_t, k_t)\rangle$, where $(x, k)$ is the instantiation of $x$ at level $k$. We say the path *enters level $k$ at $x$*, if the path contains $\langle(x, k+1), (x, k)\rangle$ as a subpath. The skip-list search procedure ensures that each level is entered at only one place, if at all. Additionally, we say the search path *enters level $k$ at partition set $S$* if it enters level $k$ at $x$ for some $x \in S$. We make the following claim, which allows for some optimization.

**Lemma 4.3** *During a $\mathrm{lookup}(y)$ operation in a B-skip-list, at each level $k$ we need never inspect an instance $(x, k)$ of element $x > y$ that lies in a different partition from the one the operation's search path entered.*

**Proof:**[sketch] At the maximum possible level $\beta$ there is nothing to prove since there is only one partition set at level $\beta$. Moreover, we find the least upper bound on $y$ in $L_\beta$. The key to the proof is to observe that once we have the least upper bound $x_{k+1}$ on $y$ in $L_{k+1}$, if we enter level $k$ at $w_k$ and proceed to search forward in $L_k$, then after coming to the final element in the partition set containing $w_k$, the very next element is $(x_{k+1}, k)$, which we have already established as being greater than $y$. This follows from the design of the partitioning. In this case we simply declare $x_{k+1}$ to be the least upper bound in $L_k$ as well as $L_{k+1}$. Otherwise, we either find $y$, in which case we are done, or we find an upper bound $x_k$ for $y$ which is smaller than $x_{k+1}$ and indeed is the least upper bound for $y$ in $L_k$. ∎

Using Lemma 4.3 it is straightforward to prove the following result.

**Lemma 4.4** *During a lookup operation, at most one partition set from each level is inspected.*

We are now ready to prove Theorem 1.2.

**Proof of Theorem 1.2:** We start by proving the expected cost bound of $O(\log_B(n))$ external memory I/Os for lookup, insert, and delete operations. Fix the capacity $N$, let $\gamma = B$, and let $\beta = \lceil \log_\gamma(N) \rceil + 2$ be the maximum level. Note the number of memory allocator operations involved in maintaining the B-skip-list partition invariant during an insertion or deletion can be amortized against the number of memory allocator operations of a lookup, up to constant factors. Lemma 4.2 then allows us to amortize – up to constant factors – the number of I/Os for an insertion or deletion against that of a lookup, and the cost of a lookup against the total number of blocks spanned by all partitions entered by the search path during the lookup. So consider a $\mathrm{lookup}(x)$ operation. By construction there are $\beta$ levels, and by Lemma 4.4 at most one partition set per level is inspected. Let $Z_i$ be the random variable equal to the number of external memory blocks spanned by the partition at level $i$ entered by the search path of the $\mathrm{lookup}(x)$ operation. The number of I/Os is then $O(\sum_{i=1}^{\beta} Z_i)$, for reasons described in the proof of Lemma 4.3. We have already shown in Lemma 4.1 that for a random partition set $S$, $\mathbf{Pr}[|S| \geq \lambda] \leq \exp(-\lambda/\gamma)$. One might be concerned that the distribution of sizes of partition sets seen on a search path to an arbitrary element might be skewed towards larger sets, and thus increase the expected number of I/Os – a data structures analogue of the "waiting time paradox" from queueing theory. We show that this is not a problem by analyzing the search path in reverse, starting at $x$ and proceeding up the levels and backward in the lists. Note that at any fixed level $i$, the search path accesses at least $z$ blocks of external memory when progressing backwards over $L_i$ only if the level $i$ partition set $S_i$ entered by the search path has at least $(z-2)B$ elements smaller than or equal to $x$. Using the same argument as in the proof of Lemma 4.1, we can prove that this occurs with probability at most $\exp(-(z-2)B/\gamma)$; essentially, we work our way backwards over $L_i$ until encountering the first element $y$ with $\mathrm{level}(y) > i$. The key point is that conditioning on the portion of the search path we have seen so far (i.e., from some element $y$ until the end the search path at $x$) does not tell us anything about the levels of elements smaller than $y$, and hence does not change the conditional distributions of their levels[3]. From this we infer

$$\mathbf{Pr}[Z_i \geq z] \leq \exp\left(-(z-2)B/\gamma\right)$$

and thus, using $\mathbf{E}[Z] = \sum_{z \geq 1} \mathbf{Pr}[Z \geq z]$ for nonnegative integral random variables $Z$, we obtain

$$\mathbf{E}[Z_i] \ \leq \ \sum_{z \geq 1} \exp\left(-(z-2)B/\gamma\right) \ = \ \frac{\exp\left(B/\gamma\right)}{1 - \exp\left(-B/\gamma\right)}$$

Linearity of expectation then yields the final bound of

$$\sum_{i=1}^{\beta} Z_i \ \leq \ \frac{\beta \cdot \exp\left(B/\gamma\right)}{1 - \exp\left(-B/\gamma\right)} \ = \ \frac{\exp\left(B/\gamma\right)}{1 - \exp\left(-B/\gamma\right)} \cdot \left(\lceil \log_\gamma(N) \rceil + 2\right).$$

Hence setting $\gamma = B$ gives a bound of $\frac{e^2}{e-1}\left(\lceil \log_B(N) \rceil + 2\right)$. The bound for range-query$(x,y)$ follows easily, as we can simply lookup$(x)$ and then scan the lowest level list $L_1$ starting from $x$ until we reach $y$, which involves reading from at most an additional $k/B + 2$ blocks of external memory, and hence $O(k/B)$ more I/Os in expectation.

We next show a bound on the space required to store the B-skip-list. The variable length key hash table which we use for memory allocation is guaranteed to require only space linear in the number of B-skip-list nodes [9]. Hence, it suffices to show that the number of B-skip-list nodes is linear in the number of list elements. Since each element $x$ is duplicated level$(x)$ times, it suffices to bound $\mathbf{E}[\text{level}(x)]$. Note that level$(x)$ is stochastically dominated by a geometric random variable with success parameter $1 - q = 1 - \frac{1}{\gamma}$, and hence has mean $1/(1-q) = \gamma/(\gamma-1)$. Additionally, by the independence of $\{\text{level}(x) : x \in \mathcal{X}\}$ we can apply standard concentration inequalities such as Chernoff bounds [6] to show that $\sum_{x \in \mathcal{X}} \text{level}(x) \leq n\gamma/(\gamma-1) + O(\sqrt{n \log n})$ with high probability. Hence the number of B-skip-list nodes is $n\gamma/(\gamma-1) + O(\sqrt{n \log n})$ with high probability, and the space required will be linear in the number of elements $n$. ∎

## 5  Conclusions

In this paper we introduced the B-skip-list, a uniquely represented data structure analogous to a B-tree which is suitable for use in uniquely represented (and hence strongly history independent) filesystems and databases. The B-skip-list supports the same operations as a B-tree, and with similar performance guarantees; it supports insertion, deletion, and lookups in $O(\log_B(n))$ external memory I/Os, where $B$ is the blocksize, supports efficient range queries, and is space efficient. The only previous uniquely represented data structure with these properties is the B-treap [10], which is significantly more complicated and more difficult to implement. Hence the principal virtue of the B-skip-list is its relative simplicity and elegance, which it inherits from the elegant skip lists of Pugh [17] and the uniquely represented hash table of Blelloch & Golovin [3, 9].

**Acknowledgements**

The author wishes to thank Guy Blelloch for helpful discussions. This work supported in part by the Center for the Mathematics of Information at the California Institute of Technology.

---

[3] Note this would not be the case if we tried to analyze the search path in the "forward" temporal direction, rather than using backwards analysis.